\documentclass{svproc}
\usepackage{graphicx}
\usepackage{url}
\usepackage {amsmath}
\usepackage{algorithmicx}
\usepackage{algpseudocode}
\usepackage{algorithm}
\usepackage{epsfig}

\begin{document}
\mainmatter              
\title{A Machine Learning - Based Migration Strategy for Virtual Network Function Instances}
\titlerunning{VNNIM}  
%
\author{Dimitrios Michael Manias, Hassan Hawilo, and Abdallah Shami}
\authorrunning{Accepted - FTC 2020} 
\institute{The Department of Electrical and Computer Engineering, Western University, Canada}

\maketitle              

\begin{abstract}
With the growing demand for data connectivity, network service providers are faced with the task of reducing their capital and operational expenses while simultaneously improving network performance and addressing the increased demand. Although Network Function Virtualization (NFV) has been identified as a promising solution, several challenges must be addressed to ensure its feasibility. In this paper, we address the Virtual Network Function (VNF) migration problem by developing the VNF Neural Network for Instance Migration (VNNIM), a migration strategy for VNF instances. The performance of VNNIM is further improved through the optimization of the learning rate hyperparameter through particle swarm optimization. Results show that the VNNIM is very effective in predicting the post-migration server exhibiting a binary accuracy of 99.07\% and a delay difference distribution that is centered around a mean of zero when compared to the optimization model. The greatest advantage of VNNIM, however, is its run-time efficiency highlighted through a run-time analysis.
\end{abstract}
\section{ Introduction}
To cope with rising capital and operational expenditures, the European Telecommunications Standards Institute (ETSI) proposed the paradigm of Network Function Virtualization (NFV) in 2012 [1]. NFV proposes the abstraction of network functions, traditionally found in proprietary hardware, and the creation of Virtual Network Functions (VNFs). These VNFs are traditionally connected to form a Service Function Chain (SFC), which, when traversed, executes a specific function. NFV is a promising technology due to its potential benefits, including faster development cycles, improved scalability, rapid reconfiguration and optimization, and a dynamic network [2]. However, there are several challenges associated with the implementation of this technology, which must be addressed before its potential benefits are realized.

One of these challenges relates to service availability. Network Service Providers (NSPs) are required to adhere to Quality of Service (QoS) guarantees and Service Level Agreements (SLA), which outline the threshold of acceptable service and network operation. When considering the various critical applications operating on these networks, including emergency and financial services, extreme care must be taken into ensuring the network is able to deliver 99.999\% availability throughout the calendar year, a number which translates to less than six minutes of downtime each year [3].

There are two main methods of improving system availability, the introduction of computational paths and live migration. A computational path can be defined as a single route that accesses VNF instances forming an SFC; however, when considering strong and resilient networks, it is common to find multiple instances of the same VNF type deployed in the network [4]. This creates multiple routes that can be traversed to deliver a service and therefore produce multiple computational paths. This idea is further illustrated below in Fig. 1, which illustrates the architecture of the virtual Evolved Packet Core (vEPC) and the idea of computational paths.

\begin{figure}
\centering
\includegraphics[width=1\textwidth]{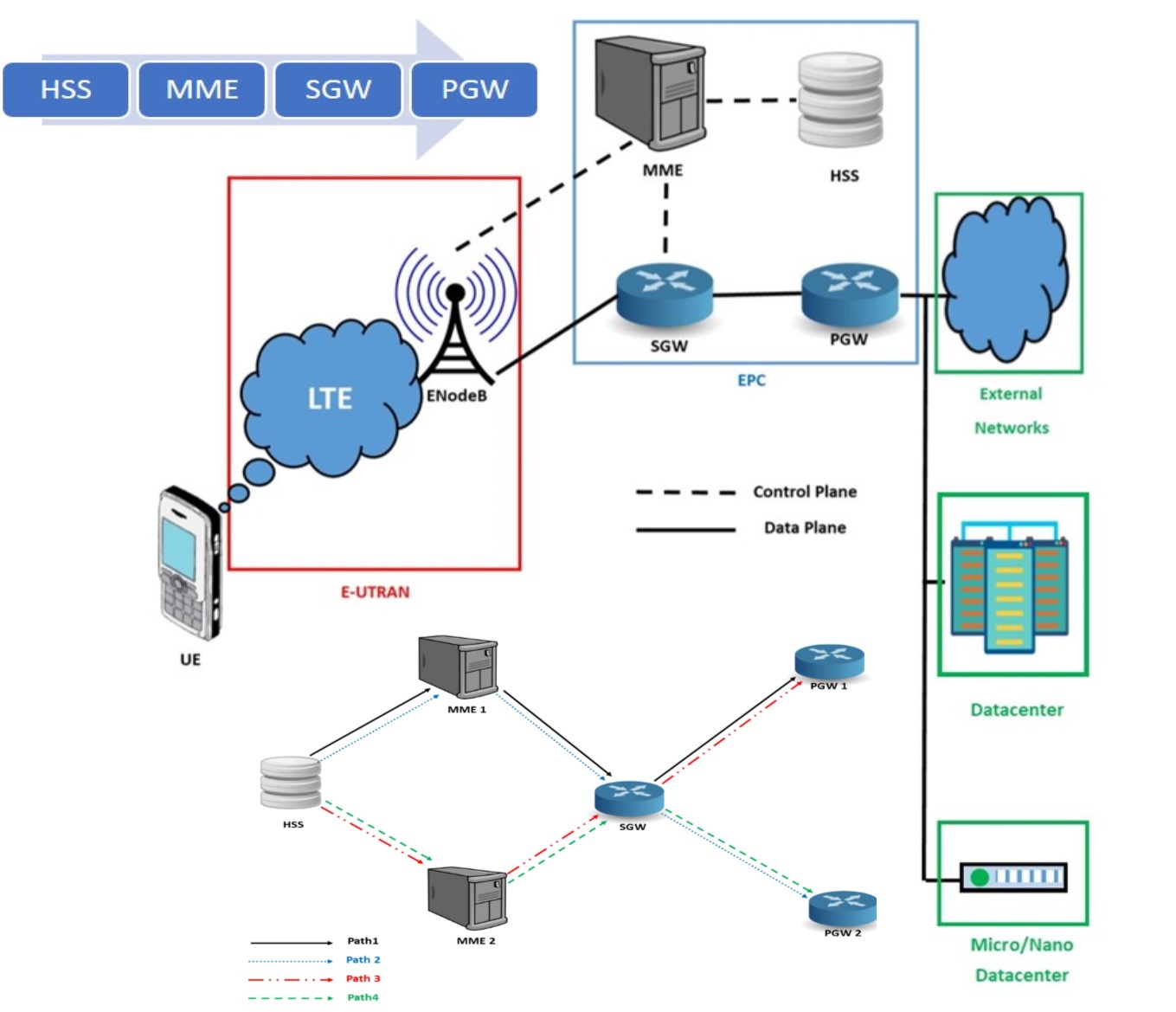}
\caption{vEPC Architecture and Computational Paths}
\end{figure}
The vEPC SFC presented in Fig. 1 is comprised of four VNF types and is the use case presented in this work; however, this work is not limited only to the vEPC use case, as it generalizes to all SFC types.

The second method of improving system availability relates to live migration, where a set of VNF instances already instantiated on network servers are migrated to different servers to improve the overall performance of the network (such as remediating performance degradation and mitigating the impact of faults). Through the consideration of the number of available computational paths, and the idea of live migration, the availability of the network can be significantly improved.

Traditionally, live migration has been conducted using optimization models or sub-optimal heuristic algorithms, which are computationally expensive and are inadequate for time-sensitive implementations that deal with critical services. To mitigate this limitation, we propose the VNF Neural Network for Instance Migration (VNNIM) migration strategy, which is capable of learning the underlying mechanisms of the optimization model and exhibits a performance that makes it feasible for real-time applications.

The remainder of this paper is structured as follows. Section II outlines related work in the field. Section III discusses the methodology used to create the VNNIM. Section IV presents and analyzes the results. Finally, Section V concludes the paper and discusses future work in the field.
\section{ Related Works}
The topic of VNF migration has been widely studied in academia. Traditionally, Integer Linear Programming (ILP), Mixed-Integer Linear Programming (MILP), and heuristic solutions have been used to solve the VNF migration problem due to its NP-hard complexity [3]. Cho {\it et al.} formulate the migration problem with the objective of minimizing the number of migrations and maximizing the reduction ratio of network latency post-migration [5]. Zhou {\it et al.} formulate the VNF migration problem with the objective of minimizing the effect of migration on the network and propose a heuristic solution considering the latency on physical links and nodes [6]. Jahromi {\it et al.} formulate an ILP optimization problem to address online VNF placement and chaining problem for value-added services in content delivery networks with the objective of minimizing network reconfiguration costs and satisfy QoS guarantees [7]. Khai {\it et al.} formulate a MILP optimization problem with the objective of minimizing the migration time as part of their flexible, multi-step approach for VM migration [8]. Hirayama {\it et al.} formulate an ILP problem to seek optimal resource allocation and train an encoder-decoder recurrent neural network, which is effective in preventing resource shortage and minimizing occurrences of VNF migration [9]. The works outlined above all address the VNF migration problem; however, they do not consider SLA requirements related to availability. Hawilo {\it et al.} propose a MILP model which considers the option of both VNF migration and re-instantiation as methods of meeting high availability SLA requirements for critical applications [3].

The work presented in this paper addresses the lack of high availability SLA requirements for critical applications while simultaneously leveraging machine learning techniques to move away from computationally intensive optimization models, thereby adopting a data-based networking paradigm. The main contributions of this paper are:

i) The introduction of the VNF Neural Network for Instance Migration (VNNIM), a model capable of predicting target servers for VNF instances selected for migration.

ii) The introduction of an optimization model used to optimize the learning rate hyperparameter to improve the performance of the VNNIM.

iii) A run-time analysis highlighting the benefits of using machine learning.
\section{ Methodology}
The following is an outline of the various steps taken to create the VNNIM, including the initial optimization model formulation, the dataset generation, the VNNIM architecture and the hyperparameter tuning. Table 1 outlines the symbols used in the Optimization Model Formulation section.

\begin{table}
\begin{center}
\caption{Optimization Model Formulation Variables}
\begin{tabular}{|c|c|}
\hline
\textbf{Symbol} & \textbf{Description}\\
\hline
{\it s} & {network server}\\
\hline
 {\it S} & {set of all network servers}\\
\hline
{$N_{s}$} & {number of network servers}\\
\hline
 {\it i} & {VNF instance}\\
\hline
 {\it I} & {set of all VNF instances}\\
\hline
 $N_{I}$ & {number of VNF instances}\\
\hline
{$I_{D}^{i}$} & {set of dependent instances for VNF instance {\it i}}\\
\hline
{$I_{M}$ }& {set of instances to be migrated}\\
\hline
{$R_{ir}$} & {computational resources {\it r} required by instance {\it i}}\\
\hline
{$R_{sr}$ }& {computational resources {\it r} available on server {\it s}}\\
\hline
{\it RE} & {set of all computational resources}\\
\hline
{$D_{i}^{P}$} & { placement delay of instance {\it i}}\\
\hline
{$D_{i}^{M}$ }& { migration overhead delay of instance {\it i}}\\
\hline
 {$D_{sd}^{ss^{*}}$}  & { delay between servers {\it s} and {\it d}}\\
\hline
{$D_{sc}^{NC}$ }&  { delay between network controller {\it c} and server {\it s}}\\
\hline
{$D_{i}^{t}$} & {delay tolerance for instance {\it i}}\\
\hline
{$D_{i}^{rec}$} & {recovery delay for instance {\it i}}\\
\hline
{$X_{is}$} & { VNF instance {\it i} is  placed on server {\it s}}\\
\hline
 {$X_{is}^{initial}$} & {VNF instance {\it i} was initially  placed on server {\it s}}\\
\hline
{$Y_{i}^{M}$} & {instance {\it i} is selected for migration}\\
\hline
\end{tabular}
\end{center}
\end{table}
\subsection{Optimization Model Formulation}
The following variables are used in the optimization problem formulation as outlined through Eq. 1-12. {\it s} denotes a network server, {\it S} denotes the set of all network servers, and $N_{s}$ denotes the number of network servers. {\it i} denotes a VNF instance, {\it I} denotes the set of all VNF instances,  $I_{D}^{i}$ denotes the set of dependent instances for instance {\it i}, $I_{M}$ denotes the set of instances to be migrated, and $N_{I}$ denotes the number of VNF instances. $R_{ir}$ denotes computational resources {\it r} required by instance {\it i}, $R_{sr}$ denotes computational resources {\it r} available on server {\it s}, and {\it RE} denotes the set of all computational resources. $D_{i}^{P}$ denotes the placement delay of instance {\it i} with respect to server delay and controller delay, $D_{i}^{M}$ denotes the migration overhead delay of instance {\it i}, $D_{sd}^{ss^{*}}$ denotes the delay between servers {\it s} and {\it d}, $D_{sc}^{NC}$ denotes the delay between network controller {\it c} and server {\it s}, $D_{i}^{t}$ denotes the delay tolerance for instance {\it i}, and $D_{i}^{rec}$ denotes the recovery delay for instance {\it i}. $X_{is}$ denotes that VNF instance {\it i} is  placed on server {\it s} , $X_{is}^{initial}$ denotes the initial VNF instance placement, and $Y_{i}^{M}$ denotes an instance is selected for migration.

The first stage in formulating the optimization problem is determining the objective function. In this work, the main objective is minimizing the downtime experienced during a migration event. The downtime experienced is defined by, the placement delay ($D_{i}^{P}$) and the migration delay ($D_{i}^{M}$).

\begin{eqnarray}
D_{i}^{P}	=X_{is}\bullet[\sum(X_{id}^{initial}\bullet D_{sd}^{ss^{*}})+D_{s}^{NC}]
\end{eqnarray}
\begin{eqnarray}
Down_{i}=D_{i}^{P}+D_{i}^{M}
\end{eqnarray}
{\it Objective:}
\begin{eqnarray}
minimize(\sum_{i}^{N_{i}}Down_{i})
\end{eqnarray}
{\it Constraints:}
\begin{eqnarray}
X_{is}=\begin{cases}
1, & \text{instance {\it i} placed on server {\it s}}\\
0, &\text{ otherwise}
\end{cases}
\end{eqnarray}
\begin{eqnarray}
Y_{i}^{M}=\begin{cases}
1, & \text{instance {\it i} selected for migration}\\
0, & otherwise
\end{cases}
\end{eqnarray}
\begin{eqnarray}
Down_{i}\text{\ensuremath{\ge}}0
\end{eqnarray}
\begin{eqnarray}
X_{is}+X_{i^{*}s}\leq2 \: \: \forall \:  i^{*} \in I_{D}^{i},\:i^{*} \in I_{M},\:D_{i^{*}}^{t}\leq D_{i}^{rec}
\end{eqnarray}
\begin{eqnarray}
X_{is}+X_{i^{*}s}^{initial}\leq2 \: \: \forall \: i^{*}\text{\ensuremath{\in}}\:I_{D}^{i},\:i^{*} \notin I_{M},\:D_{i^{*}}^{t}\leq D_{i}^{rec}
\end{eqnarray}
\begin{eqnarray}
X_{is}+X_{i^{*}s}\leq1 \: \: \forall \: i^{*}  \in I_{D}^{i},\:i^{*}  \in I_{M},\:D_{i^{*}}^{t}\geq D_{i}^{rec}
\end{eqnarray}
\begin{eqnarray}
X_{is}+X_{i^{*}s}^{initial}\leq1 \: \: \forall \:i^{*}\text{\ensuremath{\in}}\:I_{D}^{i},\:i^{*} \notin \:I_{M},\:D_{i^{*}}^{t}\geq D_{i}^{rec}
\end{eqnarray}
\begin{eqnarray}
\sum_{i}^{N_{i}}(X_{is}\text{\ensuremath{\bullet}}R_{ir})	\leq R_{sr}
\end{eqnarray}
\begin{eqnarray}
\sum_{s}^{N_{s}}X_{is}=1
\end{eqnarray}
{\it Given:}
\begin{align*}
\forall \: i \in \:I,\:s \in S,\:r \in RE,\:d \in S
\end{align*}

Eq. (4) and Eq. (5) state that the migration selection and instance placement variables must be binary. Eq. (6) states that the downtime cannot be negative. Eq. (7)-(10) are availability constraints. Eq. (11) states that a server must have enough resources to host a VNF instance. Eq. (12) states that a VNF instance can only be hosted on one server. The formulation presented above is based on the work of Hawilo {\it et al.} [3].

\subsection{ Dataset Generation}
The dataset generation was performed in two stages. The first stage involved the generation of 10,000 network snapshots with common topologies but differing network parameters as outlined in [10][11]. Each topology was comprised of 6 VNF instances and 15 servers. The initial placement of each of the 6 instances was placed using the BACON algorithm [4]. Once placed, every possible combination of instances was selected for migration ($2^{N_{I}}-1$) and evaluated using the optimization model formulated through Eq. (1 – 12). This rendered 63 differed possible migrations for each of the 10,000 network snapshots. Through solving the optimization problem 63 times per each of the 10,000 network snapshots, a total of 630,000 optimization problems were solved.

The second stage of the dataset generation involved the cleaning of the data whereby any optimization problem which did not converge to a solution was discarded from the dataset. Additionally, constant network features such as dependent VNF instance delay tolerances were dropped from the dataset as they do not contribute to the learning of the model.

\subsection{VNNIM Architecture}
The VNNIM placement model leverages machine learning through the implementation of an Artificial Neural Network (ANN). Fig. 2 outlines the architecture of VNNIM, including the number of input features ({\it F}), the number of hidden units ({\it HU}) per layer, and the number of binary outputs ({\it B}).
\begin{figure}
\centering
\includegraphics[width=1\textwidth]{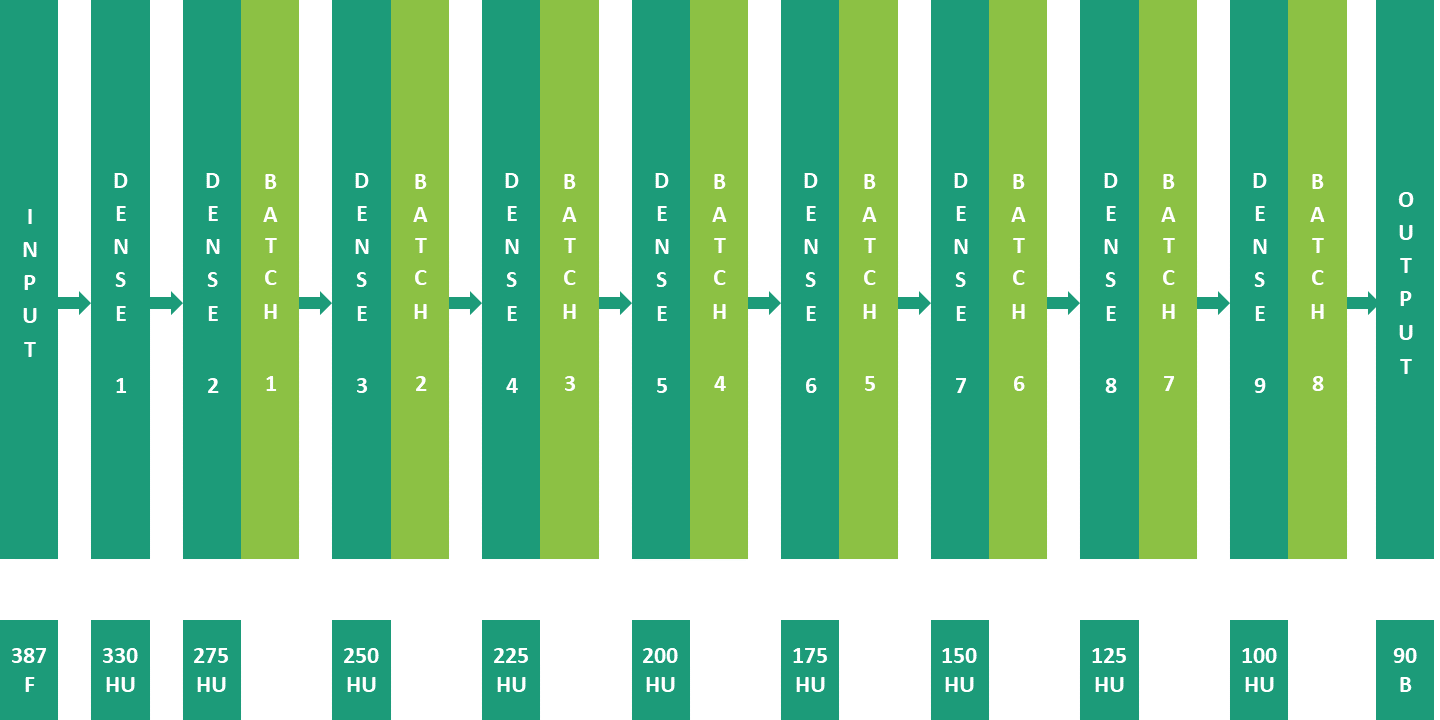}
\caption{VNNIM Architecture}
\end{figure}

There are two main types of layers present in the VNNIM architecture, the dense layer and the batch normalization layer. The dense layer is a standard, fully connected layer whereby each neuron is connected to and received input from all the neurons of the previous layer. Since neural networks are prone to overfitting and overtraining [12], normalization efforts must be used to ensure the network model learns effectively. The batch normalization layer ensures that the previous layer’s batch activations are normalized to a mean of zero and a standard deviation of 1. The current VNNIM architecture has 474,900 trainable parameters.

\subsection{Hyperparameter Optimization}
The hyperparameter optimization of the VNNIM was completed using the metaheuristic Particle Swarm Optimization (PSO) algorithm. PSO was selected because of its parallelizable nature and its ability to converge to a solution while covering a large search space. For this work, a single, tunable hyperparameter, the learning rate, was selected for optimization. The learning rate hyperparameter was selected for optimization due to its significance to the ANN training phase. In order to optimize this hyperparameter, an optimization model was formulated as follows.

Given the set of hyperparameters {\it h} and the function {\it f}, the evaluation criteria {\it E} can be formulated as a cost function. The optimization is executed across an n-fold cross-validation scheme expressed through {\it P(h)}. The hyperparameter value, which minimizes {\it P(h)}, is the optimal value.

The learning rate is the only hyperparameter selected for optimization and therefore forms the set of hyperparameters.
\begin{eqnarray}
\{ h \}=\{ learningRate \}
\end{eqnarray}
Binary cross-entropy loss ({\it bCEL}) where {\it y} represents the actual value and {\it yp} represents the predicted value was used.
\begin{eqnarray}
bCEL=-(ylog(yp)+(1-y)log(1-yp))
\end{eqnarray}
The evaluation criterion {\it E} is the loss evaluated on a model with a given learning rate {\it model(h)} across the training set {\it Ts} and validation set {\it Vs}.
\begin{eqnarray}
E(bCEL,model(h),Ts,Vs)
\end{eqnarray}
For this work, 3-fold cross-validation was used, and therefore, the final expression to be optimized is {\it P(h)}.
\begin{eqnarray}
P(h)=\frac{1}{3}\sum_{i=1}^{3}E(bCEL,model(h),Ts(i),Vs(i))
\end{eqnarray}
Finally, the objective function is formulated and constrained.
\par
{\it Objective:}
\begin{eqnarray}
minimize(P(h))
\end{eqnarray}
\par
{\it Constraint:}
\begin{eqnarray}
0.00000001\leq h\leq0.1
\end{eqnarray}

\subsection{VNNIM Training}
The VNNIM training module was completed after the dataset generation and learning rate optimization modules and is outlined as follows in Algorithm 1. For the purposes of training, the number of epochs was set to 100, the binary cross-entropy was selected as the loss function, and the binary accuracy was selected as a reporting metric.
\begin{algorithm}
\caption{VNNIM Training Process}
\label{VNNIM Training Process}
\begin{algorithmic}[1]
\State{encode categorical features using one-hot-encoding}
\State{split the training and test sets using an 80:20 scheme}
\State{normalize train and test set}
\State{$epochNumber\gets 0$}
\While{$epochNumber\le 100$}
	\State{perform neural network epoch training}
	\State{report loss and metric}
	\State{$epochNumber++$}
\EndWhile
\end{algorithmic}
\end{algorithm}

\section{Results and Analysis}
The following is a presentation of the results and an analysis of their implications on the VNNIM.
\subsection{Implementation}
Various platforms and languages were used to create and implement VNNIM. The original optimization problem formulation conducted using the DOcplex[13] Python library. Java was used to generate the various network topologies. Python was used to implement the VNNIM architecture through the Keras[14] library and to perform the hyperparameter optimization through the Optunity[15] library. All stages of the VNNIM were run on a PC with an Intel ® Core™ i7-8700 CPU @ 3.20 GHz CPU, 32 GB RAM, and an NVIDIA GeForce GX 1050 Ti GPU.
\subsection{ Dataset Generation}
The dataset generation stage of this work demonstrates a need for intelligence regarding the VNF instance migration problem. As previously outlined, the generation of 10,000 network snapshots with six instances and 15 servers resulted in 630,000 migration optimization problem solutions; however, due to the various rigid constraints imposed on the problem, the majority of the optimizations did not converge to a solution (unsuccessful optimization) and therefore failed to produce usable data. Approximately 36\% of the optimizations were successful, significantly reducing the expected dataset of 630,000 points to 226,360. While there is still a significant amount of data remaining, the distribution of the resulting dataset must be examined.

Fig. 3 shows the ratio of successful versus unsuccessful optimizations with respect to the number of instances migrated.
\begin{figure}
\centering
\includegraphics[width=1\textwidth]{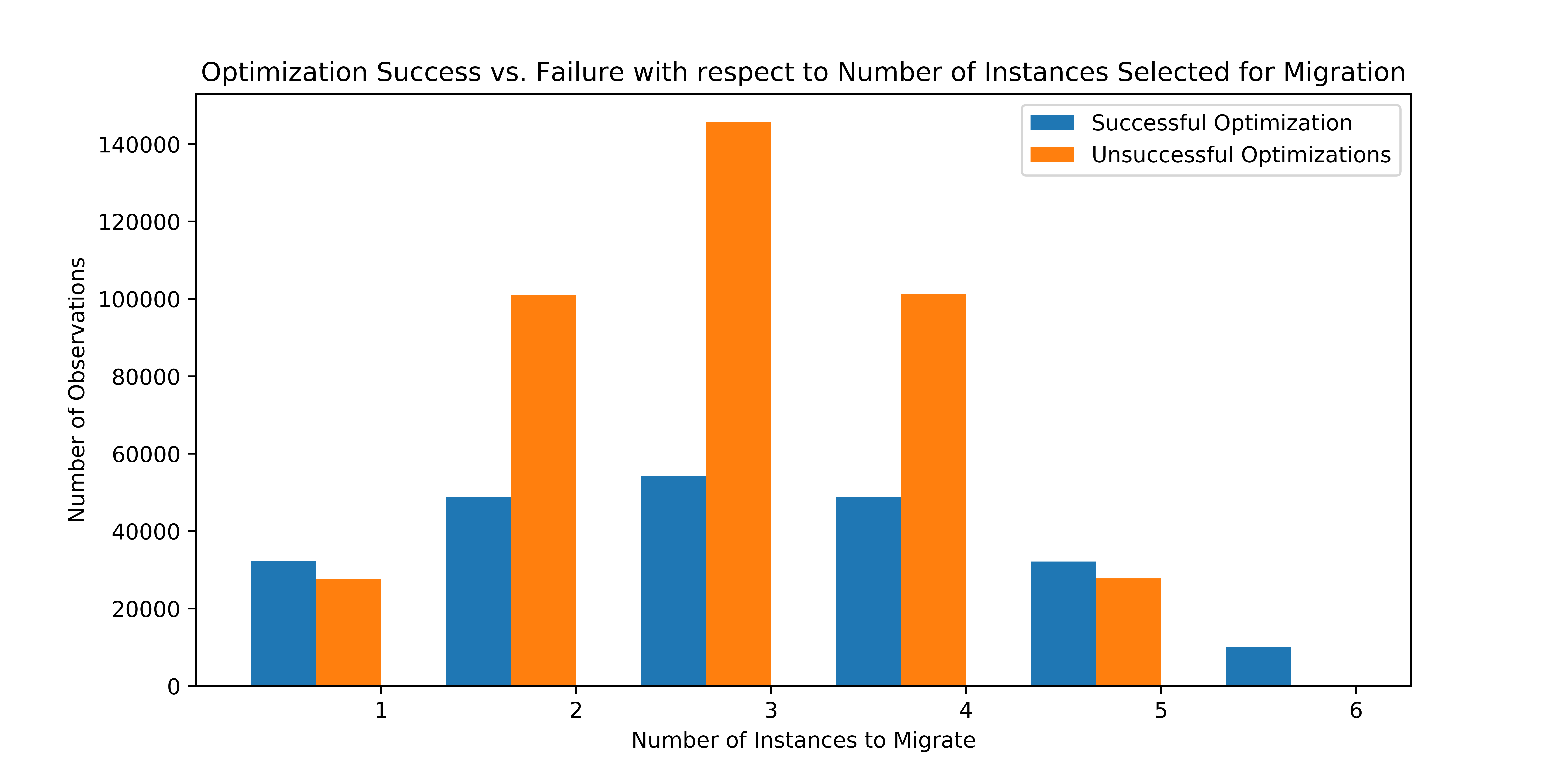}
\caption{Optimization Success vs. Number of Instances to Migrate}
\end{figure}

As seen in this figure, there is a significant imbalance in the dataset, especially when considering the scenario when all six instances require migration. Additionally, it is presumed that as the number of instances requiring migration increases, so too does the complexity of the server prediction. Ideally, there would be an equal distribution of data or; there would be more data available as the number of instances to migrate increases. The optimization model fails to converge to a solution most commonly when the number of instances to migrate ranges from two to four. Additionally, a key observation is that there are zero unsuccessful optimizations when all instances are selected for migration. This is an important observation as it means that through the generation of additional network snapshots and considering only the case where all six instances require migration, the dataset can be increased, and the distribution further balanced.
\subsection{Hyperparameter Optimization}
The results of the hyperparameter optimization resulted in an optimal learning rate of 0.002318. This value was determined by running the optimization model outlined in Eq. (13-18) for 50 iterations. The results of the optimization are displayed in Fig. 4, where the binary cross-entropy loss is expressed as a function of the learning rate.
\begin{figure}

\centering
\includegraphics[width=1\textwidth]{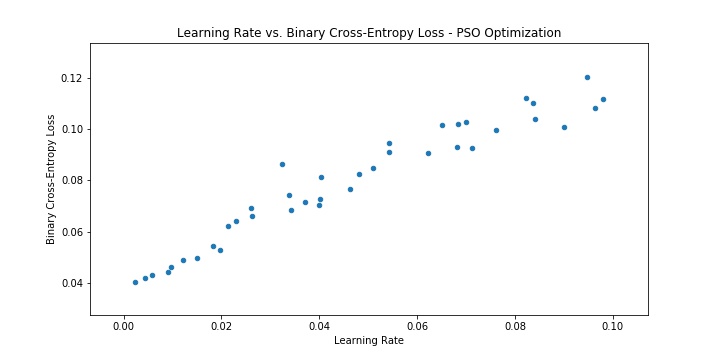}
\caption{Learning Rate vs. Binary Cross-Entropy Loss}
\end{figure}
From the Fig. 4, it is evident that as the learning rate is decreased, the loss function decreases. Given the size of the search space and the number of iterations completed, the optimal value of the learning rate forms a local minimum as the entirety of the search space has not been explored. The goal of this hyperparameter optimization was to determine the relationship between the learning rate and the loss; as an extension to this work, the learning rate search space will be reduced to determine where the reduction of the learning rate results in a plateau in terms of the minimization of the loss function and the global minimum will be determined.
\subsection{VNNIM Performance}
There are two methods used to evaluate the performance of the VNNIM migration strategy, the predictive accuracy of the model and the observed delay across all computational paths. These two methods were selected because they provide both a high-level overview of the effectiveness of the model to learn (accuracy) and a domain-based evaluation of its efficiency when the optimal placement has not been learned. This analysis will be one of the main criteria for evaluating the feasibility of this migration strategy.
\subsubsection{Accuracy Evaluation}
When evaluating the accuracy of the VNNIM model, there are several metrics to consider. Firstly, since the output of the neural network is a binary array, the binary predictive accuracy must be considered. This accuracy is a critical indicator as to how the model is performing given the nature of the prediction; for a model to be effective in learning from the data provided, it must learn the critical relationship that each instance must have only one associated server. To this end, 6/90 predictions must be “1”, and 84/90 predictions must be “0”. This overwhelming majority of “0” predictions required can be problematic for the VNNIM model during training since underfitting the data and predicting an output of all “0” would result in a respectable binary accuracy of 93.33\%. Proper fitting of the model would result in a binary accuracy greater than 93.33\% and ideally approaching 100\%. The binary accuracies throughout the training and testing phases were 99.11\% and 99.08\%, respectively.

Binary accuracy, however, is not the only metric that should be used to evaluate the performance of the model; the categorical accuracy should also be considered. The categorical accuracy is defined as the number of correctly predicted servers. This metric is more indicative of the model’s performance as its main objective is to accurately predict the destination server for an instance which has been selected for migration. To gain further insight into model performance, the categorical accuracy was calculated with respect to the number of migrated instances, as observed in Fig. 5. While the binary and categorical accuracy metrics give significant insights into model performance, the overall performance of the model must be judged based on the complete predictive accuracy of the system. For this metric, if the entire prediction is correct ({\textit i.e.} all six instances are placed on the optimal server), it receives a binary value of “1”; otherwise, a binary value of “0” is assigned. When evaluating this metric with respect to the number of instances selected for migration, the distribution, also presented in Fig. 5 is observed.
\begin{figure}

\centering
\includegraphics[width=1\textwidth]{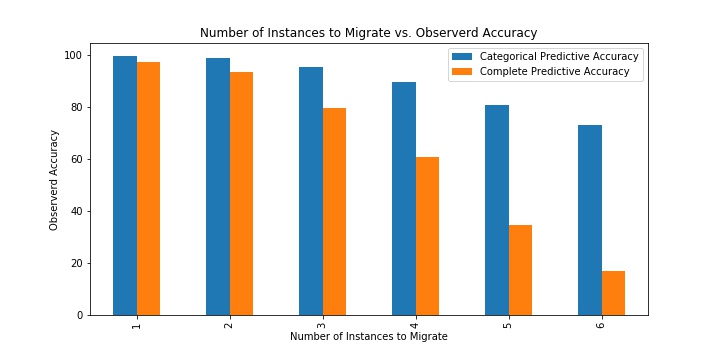}
\caption{Observed Categorical Accuracy}
\end{figure}
As seen in Fig. 5, the VNNIM exhibits outstanding performance in terms of categorical accuracy when a single instance or a few instances are selected for migration and its performance beings to drop as the number of instances selected for migration increases. This is an expected outcome due to the increasing complexity of the system as the number of instances selected for migration increases coupled with the disproportionality in data previously defined. Additionally, the VNNIM model exhibits excellent complete predictive accuracy when the number of instances to migrate is low; however, this accuracy drops drastically as the number of instances requiring migration increases. Additionally, the common cases where a few instances are selected for migration exhibits very good complete predictive accuracy, indicating VNNIM is performing well. Furthermore, it can be observed that a small change in the categorical predictive accuracy translates to a significant change in the complete predictive accuracy. Finally, while there is a decrease in complete predictive accuracy, this cannot be quantified without evaluating the model using a domain-based metric such as the delay exhibited between interdependent VNF instances and across the SFCs.
\subsubsection{Delay Evaluation}
In order to fully appreciate the performance of VNNIM, a probability density function of the delay difference across all computational paths was generated, as exhibited in Fig. 6.
\begin{figure}

\centering
\includegraphics[width=0.75\textwidth]{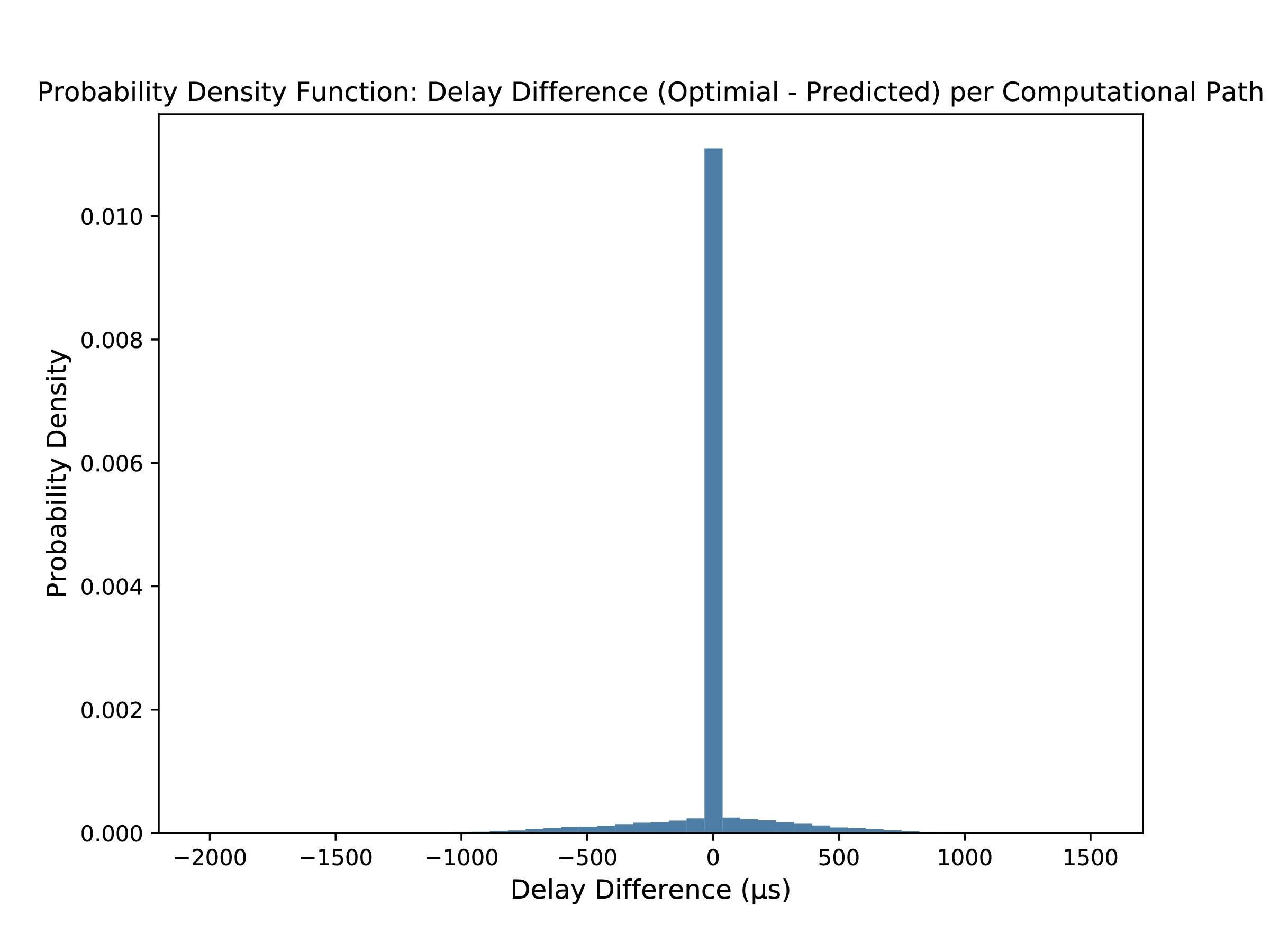}
\caption{Probability Density Function Delay Difference}
\end{figure}

This distribution was calculated by subtracting the delay experienced by the VNNIM prediction placements from the delay experienced by the optimization model placements across all four computational paths per each trial in the test set. Fig. 6 clearly illustrates the effectiveness of VNNIM as the distribution is extremely narrow, with a spike at a mean of 0. Additionally, this distribution has minor tails, indicating that the standard deviation is not significant. By observing the tails, it is evident that the positive and negative values tend to cancel out as the tails are nearly symmetrical. This symmetry suggests that in the event the VNNIM model does not accurately predict the placement of the post-migration instance, the delay difference experienced across all resulting computational paths is almost zero. Fig. 7 further supports this as it shows the difference in delay between dependent VNF instances when using VNNIM and the optimal solution. 

When considering Fig. 7, it must be noted that since we measure the delay observed between dependent instances, there will be paths in which the optimal solution appears to produce a larger delay than the VNNIM however, when considered across the entire service function and across all computational paths, the optimal solution produces a slightly better end-to-end delay.
\begin{figure}

\centering
\includegraphics[width=1\textwidth]{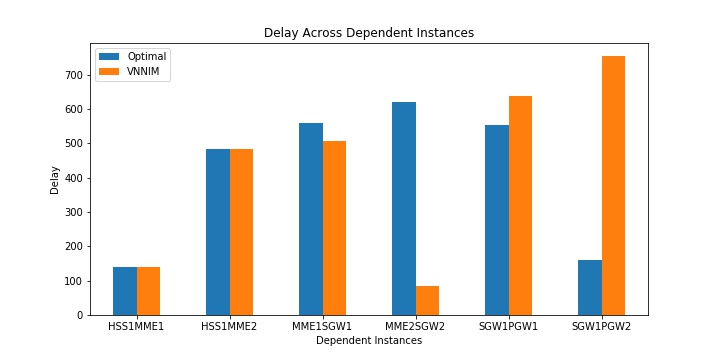}
\caption{Delay Across Dependent Instances}
\end{figure}
As seen in Fig. 7, in the case where the same server has not been selected, the delay between dependent instance delays varies ({\textit i.e.} Sometime VNNIM produces less delay, other times the optimal solution does). However, when considering the delay experienced across all computational paths, the delay differences average to approximately zero.
\subsection{Run-time Analysis}
Perhaps the greatest advantage of the VNNIM model is its run-time characteristics. When considering optimization problems, specifically the VNF migration problem, which has been defined as NP-hard, a major limitation is exhibited through their inadequacy for real-time implementations. Since the computationally intensive phase of VNNIM, the training and hyperparameter optimization is completed offline; its real-time execution is exceptionally efficient as it involves only the prediction phase. Table 2  lists some run-time characteristics observed between the optimization model and VNNIM when handling multiple migration requests.
\begin{table}
\begin{center}
\caption{Run-time Analysis}
\begin{tabular}{c|c|c}
\textbf{Number of Migration Requests} & \textbf{Optimization Time} & \textbf{VNNIM Time}\\
\hline
1 & 1.4273 {\it s} & 0.0087 {\it s} \\
10 & 10.6233 {\it s} & 0.0106 {\it s} \\
100 & 130.6602 {\it s} & 0.0362 {\it s} \\
1000 & 1147.3395{\it s} & 0.2210 {\it s} 
\end{tabular}
\end{center}
\end{table}
Table 2 shows that as the number of migration requests increases, the time required by the optimization model increases proportionally. However, when considering the VNNIM, it can be seen that the amount of time taken to perform a prediction is significantly less. This run-time analysis demonstrates a fundamental need for the VNNIM, and the continual improvement of the model will produce a fully implementable, real-time solution to the VNF migration problem.
\section{Conclusions and Future Work}
The work presented above introduces the VNF Neural Network for Instance Migration (VNNIM), a model capable of learning the underlying relationships of network features and effectively predicting the target server for a VNF instance migration. Aside from its excellent performance when a single, or a couple of instances are selected for migration, the VNNIM excels due to its ability to predict target servers in real-time, something which was previously unattainable using only complex optimization models. Future work in this field will involve the expansion of the hyperparameter search space to include additional hyperparameters that currently have not been optimized. Additionally, a change in the VNNIM architecture will see the addition of dropout layers to prevent overfitting during longer training sessions. Furthermore, feature dimensionality reduction will be considered in addition to the expansion of this model to address larger networks.

%
%

\end{document}